\documentclass[twoside,twocolumn,9pt]{article}
\usepackage{extsizes}
\usepackage[super,sort&compress,comma]{natbib} 
\usepackage[version=3]{mhchem}
\usepackage[left=1.5cm, right=1.5cm, top=1.785cm, bottom=2.0cm]{geometry}
\usepackage{balance}
\usepackage{mathptmx}
\usepackage{sectsty}
\usepackage{graphicx} 
\usepackage{lastpage}
\usepackage[format=plain,justification=justified,singlelinecheck=false,font={stretch=1.125,small,sf},labelfont=bf,labelsep=space]{caption}
\usepackage{float}
\usepackage{fancyhdr}
\usepackage{fnpos}
\usepackage[english]{babel}
\addto{\captionsenglish}{%
  
}
\usepackage{array}
\usepackage{droidsans}
\usepackage{charter}
\usepackage[T1]{fontenc}
\usepackage[usenames,dvipsnames]{xcolor}
\usepackage{setspace}
\usepackage[compact]{titlesec}
\usepackage{hyperref}
\usepackage[usenames,dvipsnames]{xcolor}

\definecolor{cream}{RGB}{222,217,201}

\usepackage{epstopdf}%This line makes .eps figures into .pdf - please comment out if not required.

\definecolor{cream}{RGB}{222,217,201}

\begin{document}

\pagestyle{fancy}
\thispagestyle{plain}
\fancypagestyle{plain}{
%%%HEADER%%%
\renewcommand{\headrulewidth}{0pt}
}
%%%END OF HEADER%%%

%%%PAGE SETUP - Please do not change any commands within this section%%%
\makeFNbottom
\makeatletter
\renewcommand\LARGE{\@setfontsize\LARGE{15pt}{17}}
\renewcommand\Large{\@setfontsize\Large{12pt}{14}}
\renewcommand\large{\@setfontsize\large{10pt}{12}}
\renewcommand\footnotesize{\@setfontsize\footnotesize{7pt}{10}}
\makeatother

\renewcommand{\thefootnote}{\fnsymbol{footnote}}
\renewcommand\footnoterule{\vspace*{1pt}% 
\color{cream}\hrule width 3.5in height 0.4pt \color{black}\vspace*{5pt}} 
\setcounter{secnumdepth}{5}

\makeatletter 
\renewcommand\@biblabel[1]{#1}            
\renewcommand\@makefntext[1]% 
{\noindent\makebox[0pt][r]{\@thefnmark\,}#1}
\makeatother 
\renewcommand{\figurename}{\small{Fig.}~}
\sectionfont{\sffamily\Large}
\subsectionfont{\normalsize}
\subsubsectionfont{\bf}
\setstretch{1.125} %In particular, please do not alter this line.
\setlength{\skip\footins}{0.8cm}
\setlength{\footnotesep}{0.25cm}
\setlength{\jot}{10pt}
\titlespacing*{\section}{0pt}{4pt}{4pt}
\titlespacing*{\subsection}{0pt}{15pt}{1pt}
%%%END OF PAGE SETUP%%%

%%%FOOTER%%%
\fancyfoot{}
\fancyfoot[LO,RE]{\vspace{-7.1pt}\includegraphics[height=9pt]{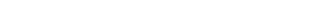}}
\fancyfoot[CO]{\vspace{-7.1pt}\hspace{13.2cm}\includegraphics{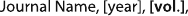}}
\fancyfoot[CE]{\vspace{-7.2pt}\hspace{-14.2cm}\includegraphics{RF}}
\fancyfoot[RO]{\footnotesize{\sffamily{1--\pageref{LastPage} ~\textbar  \hspace{2pt}\thepage}}}
\fancyfoot[LE]{\footnotesize{\sffamily{\thepage~\textbar\hspace{3.45cm} 1--\pageref{LastPage}}}}
\fancyhead{}
\renewcommand{\headrulewidth}{0pt} 
\renewcommand{\footrulewidth}{0pt}
\setlength{\arrayrulewidth}{1pt}
\setlength{\columnsep}{6.5mm}
\setlength\bibsep{1pt}
%%%END OF FOOTER%%%

%%%FIGURE SETUP - please do not change any commands within this section%%%
\makeatletter 
\newlength{\figrulesep} 
\setlength{\figrulesep}{0.5\textfloatsep} 

\newcommand{\topfigrule}{\vspace*{-1pt}% 
\noindent{\color{cream}\rule[-\figrulesep]{\columnwidth}{1.5pt}} }

\newcommand{\botfigrule}{\vspace*{-2pt}% 
\noindent{\color{cream}\rule[\figrulesep]{\columnwidth}{1.5pt}} }

\newcommand{\dblfigrule}{\vspace*{-1pt}% 
\noindent{\color{cream}\rule[-\figrulesep]{\textwidth}{1.5pt}} }

\makeatother
%%%END OF FIGURE SETUP%%%

%%%TITLE, AUTHORS AND ABSTRACT%%%
\twocolumn[
  \begin{@twocolumnfalse}
{\includegraphics[height=30pt]{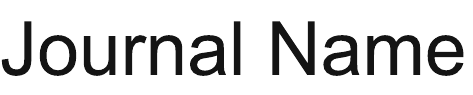}\hfill\raisebox{0pt}[0pt][0pt]{\includegraphics[height=55pt]{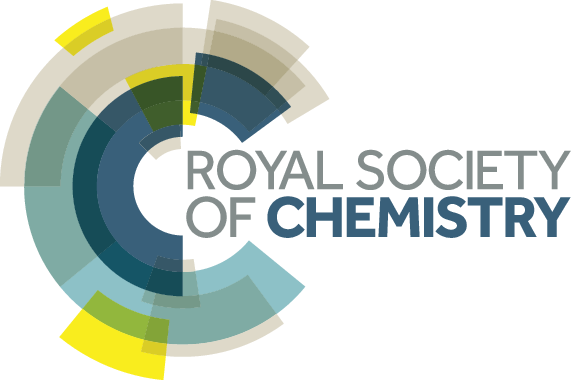}}\\[1ex]
\includegraphics[width=18.5cm]{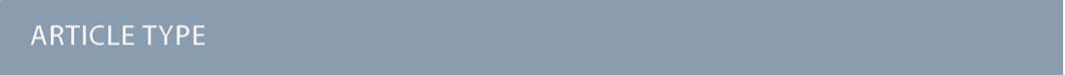}}\par
\vspace{1em}
\sffamily
\begin{tabular}{m{4.5cm} p{13.5cm} }

\includegraphics{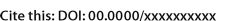} & \noindent\LARGE{\textbf{Continuous alloying between rocksalt and half-Heusler structures drives metal-semiconductor transition in ErNi$_x$Sb}} \\

%and suppression of phonon propagation

\vspace{0.3cm} & \vspace{0.3cm} \\
& \noindent\large{Maria Wróblewska\textit{$^{a, b}$}, Eric S. Toberer\textit{$^{a}$}, and Kamil M. Ciesielski$^{\ast}$\textit{$^{a}$}  }\\

\includegraphics{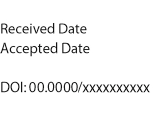} & \noindent\normalsize{\textit{XYZ} half-Heusler phases are often described as a \textit{XZ} rock salt sublattice with an interstitial \textit{Y} atom. 
However, transport properties across a solid solution between rock salt and half-Heusler structures has not previously been studied. 
In this paper, we demonstrate the exceptional tolerance of Ni vacancies in ErNi$_x$Sb, resulting in a complete alloy in the ErSb-ErNiSb space. 
Thermoelectric characterization demonstrates a continuous electronic transition associated with the gradual collapse of the band gap with Ni removal.  
The carrier concentration increase by three orders of magnitude and Seebeck coefficient decreases from 260 $\mu$V/K to $<$5 $\mu$V/K. 
Speed  of sound measurements indicate that removal of Ni softens the lattice, consistent with the breakdown of the covalent Ni-Sb sublattice. 
For low Ni content compositions, the combination of low speed of sound coupled with Ni vacancies strongly hamper  the propagation of phonons. 
The ability to control the electronic and thermal transport properties across the rock salt to half-Heusler continuum 
opens new material design strategies across thermoelectricity, superconductivity, and non-trivial topology.
} \\
\end{tabular}
 \end{@twocolumnfalse} \vspace{0.6cm}
  ]

\renewcommand*\rmdefault{bch}\normalfont\upshape
\rmfamily
\section*{}
\vspace{-1cm}

\footnotetext{\textit{$^{a}$}~Department of Physics, Colorado School of Mines, Golden, Colorado 80401, United States.}
\footnotetext{\textit{$^{b}$}~Faculty of Chemistry, Warsaw University of Technology, Warsaw, Poland.}
\footnotetext{\textit{$^{\ast}$} E-mail: kciesielski@mines.edu}

\footnotetext{\dag~Electronic Supplementary Information (ESI) available. See DOI: 00.0000/00000000.}

\section{Introduction}
\begin{figure}[t]
 \centering
   \includegraphics[width=\columnwidth]{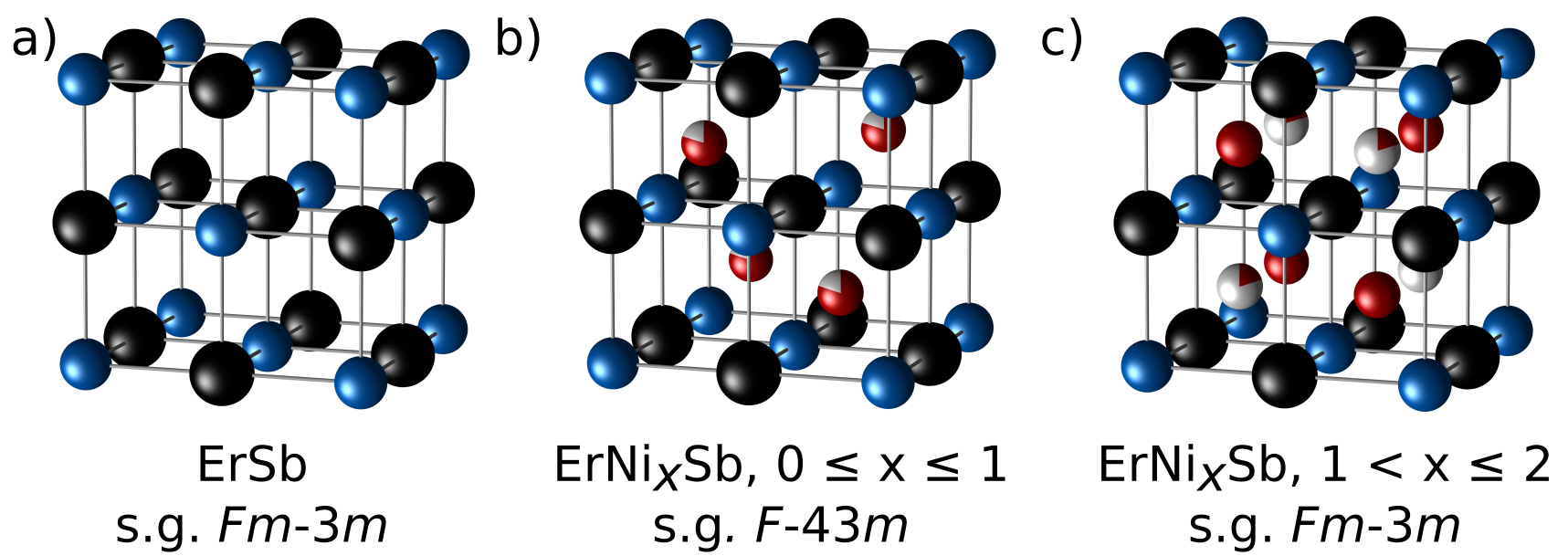}
   \caption{Crystal structures of the examined chemical system: a) rocksalt, b) half-Heusler structure, c) full-Heusler-like structure.}
    \label{fig:structures}
 \end{figure}

Half-Heusler (HH) phases are ternary intermetallic compounds with general formula $XYZ$; they are composed of group I-IV elements ($X$), late transition metals ($Y$), and $p$-block elements ($Z$).
Materials from this group crystallize in the MgAgAs-type unit cell, s.g. $F$-43$m$ (Figure \ref{fig:structures}b) and are notorious for crystallographic disorder. It includes the proclivity towards the formation of intrinsic antisite and interstitial defects\cite{xie2014intrinsic, xie2012interrelation}, phase separation\cite{rausch2015fine, page2016origins},  occurrence of vacancy short-range order\cite{xia2019short, rausch2016short} and split atomic positions\cite{dong2022half, ciesielski2020structural, synoradzki2019thermal}. 
Largely unexplored, however, are HH's ability to transition between the adjoining rock salt and full Heusler crystal structures (Figure \ref{fig:structures}) via solid solutions\cite{mishraa2002ternary, dong2022half}.
In this paper, we focus on the rock salt to Half Heusler solid solution and the evolution of elastic, electronic, and thermal properties therein.

Largely unexplored, however, are HH's ability to transition between the adjoining rock salt and full Heusler crystal structures  via solid solutions\cite{mishraa2002ternary, dong2022half}.
Figure \ref{fig:structures} highlights the structural relationships between these crystal structures. 
In this paper, we focus on the rock salt to Half-Heusler part of this space, as well as the resulting evolution of elastic, electronic, and thermal properties.

HHs exhibit a vast array of properties `on request' arising from both chemical diversity and structural complexity.
The intriguing behaviors for which HH were studied recently include  superconductivity (e.g. ErPdBi)\cite{radmanesh2018evidence, nakajima2015topological, tafti2013superconductivity}, that can coexist with long-range magnetic order\cite{nikitin2015magnetic,pavlosiuk2016antiferromagnetism,  pan2013superconductivity} as well as topological quantum effects (e.g. LuPtBi) \cite{lin2010half, hirschberger2016chiral,  chadov2010tunable, suzuki2016large, shekhar2018anomalous,  xiao2010half}. 
From the applied perspective, HH phases are also excellent thermoelectrics (e.g. NbFeSb, (Zr,Ti,Hf)NiSn,  ZrCoBi)\cite{quinn2021advances, zeier2016engineering, zhu2015high, fu2015realizing, li2024half}. Compounds from this group are known for their exceptional electronic contribution to thermoelectric performance, which largely stems from their high symmetry crystal structure and large band degeneracy\cite{fu2014high, zhu2018discovery}. On the other hand, the high symmetry of HH crystal structure paired with relatively stiff bonding leads to high lattice thermal conductivity\cite{ren2023vacancy, zeier2016engineering}, which is detrimental from a thermoelectric point of view. Efforts to reduce the lattice thermal conductivity have included nanostructuring and isoelectronic alloying\cite{quinn2021advances, zeier2016engineering, zhu2015high, fu2015realizing, li2024half}. 

Beyond alloying within the HH structure, recent work has considered alloying between distinct structure types (Figure \ref{fig:structures}).  TiRu$_x$Sb ($x$ = 1.15-2) was recently reported with homogeneity range spanning between HH-like and full-Heusler  structures\cite{dong2022half}. The disorder allowed band structure evolution including switching between $n$-type and $p$-type transport regimes and led to reduction in phonon propagation. 
Furthermore, YbNiSb was reported with homogeneity range spanning between rocksalt YbSb and HH YbNiSb\cite{mishraa2002ternary}. Transport properties were not studied for this curious material. While the above scenario discusses \textit{XZ-XYZ} solid solutions, we are not aware of alloys between zincblende (\textit{YZ}) and HH structure.

In this project we focus on alloy system ErNi$_x$Sb as an analog to YbNiSb; Erbium is less expensive, and easier to work with than the volatile Yb. ErNiSb has also shown some preliminary potential in thermoelectric evaluations\cite{kawano2008substitution, ciesielski2020thermoelectric, ciesielski2020high}. Theoretical calculations for HH ErNiSb predict  a finite, narrow band indirect gap of ca. 0.3 eV. With removal of nickel (\textit{i.e.} for ErSb) the bands are expected to collapse and form a metal\cite{lukoyanov2022ab}. 
In ErSb, the overlap between the former conduction and valence bands is \textit{ca.} 1.5 eV.
Recently, \textit{ab-initio} defect energy calculations shown low vacancy formation energy on Ni site in ErNiSb and other half-Heuslers\cite{chen2024ni}. 
Ni vacancies were found to be charge-neutral defects throughout the band gap, consistently with formal zero valence of Ni in $d^{10}$ state. %The low energy formation of vacancies was ascribed therein to the weak bonding between Er and Ni. 

Here, we experimentally investigate the existence of the  ErNi$_x$Sb ($0\leq x \leq1.2$) solid solution and the influence of Ni occupation on elastic, electronic and thermal properties. 
Bulk pellets are prepared via arc melting and hot pressing and solubility is interrogated with diffraction and electron microscopy.  
Changes in bonding character are assessed based on sound velocity measurements. 
High temperature thermoelectric transport measurements include Hall and Seebeck coefficients, as well as electrical resistivity. 
For semiconducting compositions, the transport properties are analyzed within a single parabolic band approximation.  
Lattice thermal conductivity is studied to assess the impact of Ni vacancies on phonon transport.  
Ultimately, we find that a full solid solution exists in ErSb-ErNiSb space; traversing this continuum induces both a metal-semiconductor transition and a softening of the crystal lattice.  
Within the alloy, abundant vacancies turn out to be very efficient phonon scattering centers.

\begin{figure} [t]
\centering
  \includegraphics[width=\columnwidth]{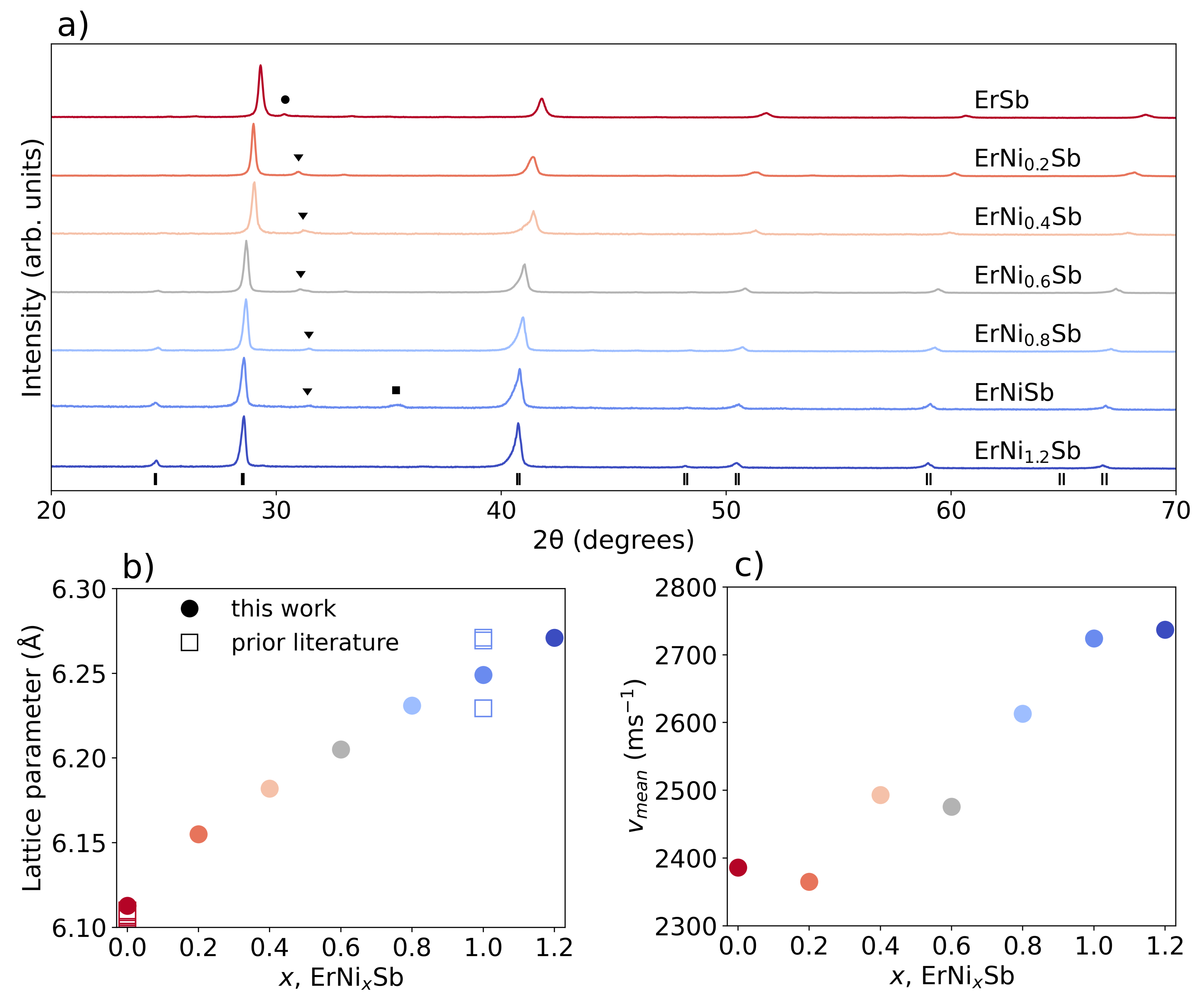}
  \caption{a) X-ray powder diffraction patterns for ErNi$_x$Sb samples. Bragg maxima shifting towards lower angles denote shrinkage of the unit cell with removal of nickel. Small amounts of impurities are represented by circle (Er\textsubscript{2}O\textsubscript{3}), triangle (NiSb) and the square (plausible Er$_3$Ni$_6$Sb$_5$). b)~Lattice parameter obtained from Rietveld refinement prove existence of solid solution in ErNi$_x$Sb space. Literature lattice parameters are taken from Refs.\cite{a_ErSb1, a_ErSb2, a_ErSb3, a_ErSb4, a_ErSb5, a_ErSb6, a_ErSb7, a_ErNiSb1, a_ErNiSb2, a_ErNiSb3, a_ErNiSb4}. c) Sounds velocity increases with rising nickel content in ErNi$_x$Sb, which indicates changing bonding character from ionic to mixed covalent-ionic.}
  \label{fig:XRDandStuff}
\end{figure}

\section{Methods}

The synthesis was performed via combination of arc-melting and hot-pressing. The samples from series ErNi$_x$Sb, $x$ = 0, 0.2, 0.4, 0.6, 0.8, 1.0, 1.2 were arc-melted under ultra high-purity (5N) argon using elements elements: Er chunk 99.9 at.\%, Ni wire 99.99 at.\%, Sb shot 99.999 at.\%.
Evaporation of antimony during the melt was compensated by adding extra Sb in a quantity 5-10\% of the sample weight beforehand the synthesis.  Each specimen was flipped over and remelted at least 3 times to ensure homogeneity, until Sb excess was removed.
Afterwards, the ingots were hand-grounded in agate mortar and hot pressed in temperature range 1000-1150 $^{\mathrm{o}}$C, 50 MPa, 15-180 min, see details in Table S1. Higher temperature and pressing time was generally necessary to obtain dense pellets with compositions having lower Ni content.
The resultant density calculated with geometrical method was always higher than 93\%.

X-ray diffraction was performed on Bruker D2 Phaser device using Cu K$\alpha$ radiation source. 
Rietveld analysis was carried out with Topas Academic v7 software
Microstructure of the pellets was studied using FEI Quanta 600i Environmental Scanning Electron Microscope (SEM), employed with energy dispersive spectroscopy (EDS). Temperature dependencies of electrical resistivity and Hall effect were studied with the Van der Pauw geometry on a custom built apparatus\cite{borup2012measurement}. The device uses electromagnet with a field of 1 T; the current supplied for those measurements was 250 mA. Measurements of the thermopower were performed on a custom built device\cite{iwanaga2011high}. High-temperature cycles of heating and cooling for both Hall effect and Seebeck coefficient measurements were performed to ensure that samples show no thermal hysteresis. Sound velocity studies at room temperature were performed with a pulse-echo transducer setup using longitudinal and transverse transducers (Olympus 5072PR Pulser/Receiver). Thermal diffusivity was measured using a commercial Netzsch LFA 467 flash diffusivity system. Graphite spray coating on specimens was employed for the diffusivity measurements to ensure proper absorption of the laser flash and suppress parasitic emissivity. The thermal conductivity was obtained with formula $\kappa$ = $Dd_{exp}C_{p}$, where $D$ stands for the diffusivity, $d_{exp}$ denotes the density obtained with the geometrical method, and $C_p$ is heat capacity calculated from Dulong-Petit limit.

\section{Results and discussion}

\subsection{Crystal structure and microstructure}

Diffraction patterns for all the pellets are shown on Fig.~\ref{fig:XRDandStuff}a.  ErSb was obtained as nearly single-phase sample crystallizing in the rocksalt structure, s.g. $Fm$-$3m$, \textit{c.f.} Fig. \ref{fig:structures}a. Adding nickel into the system results in shifting of the Bragg maxima towards lower angles, which indicates expanding of the unit cell. The observation can be explained by filling tetrahedral voids in rocksalt cell, \textit{i.e.} the gradual formation of the half-Heusler structure, s.g. $F$-$43m$, see Fig. \ref{fig:structures}b. Lattice parameter ($a$) values resultant from Rietveld refinement are displayed in Fig. \ref{fig:XRDandStuff}b. For ErSb, the obtained value of \textit{a} is in line with the previous literature\cite{a_ErSb1, a_ErSb2, a_ErSb3, a_ErSb4, a_ErSb5, a_ErSb6, a_ErSb7, a_ErNiSb3}.
In the case of ErNiSb, the prior literature exhibits significant discrepancy in values of $a$ in the range $6.229-6.271$~\AA \cite{a_ErNiSb1, a_ErNiSb2, a_ErNiSb3, a_ErNiSb4}. Our result for sample with $x$ = 1 is somewhat in the middle of the reported range.
For the remaining ErNi$_x$Sb samples, the values of lattice parameter change according to Vegard's law in range of $x$ = 0 - 1.2, demonstrating solid solubility between the rocksalt and half-Heusler structures.  The studied samples contain small amount of either Er$_2$O$_3$, or Ni-rich impurities: NiSb, and the plausible Er$_3$Ni$_6$Sb$_5$ phase (see the structural prototype Y$_3$Ni$_6$Sb$_5$, ref.\cite{stoyko2014crystal}). We anticipate that the severe disorder in ErNi$_x$Sb could potentially lead to some interesting short-range atomic arrangements. Hence,  structural studies with more intricate techniques appear as promising direction for the future research.

Electron microscopy shown that all the samples are composed of mostly of ErNi$_x$Sb as a main phase, see Fig. S2. Qualitatively, composition of the main phase was in all cases slightly depleted in nickel with respect to nominal compositions. Consistent with this observation, we were able to detect Ni-rich impurities in most of the samples. ErSb specimen, interestingly, had small amount of precipitation with approximate ErSb$_{0.67}$ composition, which can correspond to Er$_5$Sb$_3$ phase. This phase could give similar signal in XRD to Er$_2$O$_3$, when present in small amounts.

\subsection{Chemical bonding}

As a next step we will briefly consider changes in chemical bonding, based on a simple macroscopic indicator: speed of sound.
Materials from HH group are known to exhibit stiff bonding that leads to high speeds of sound (3000-4000 m/s), see Ref.\cite{ren2023vacancy}. From a thermoelectric perspective, low speed of sound and related reduced thermal conductivity are beneficial. In our case, the sound velocity for ErNiSb,  Fig. \ref{fig:XRDandStuff}c, is only 2750 m/s.  
Several recent materials from HH group have also exhibited significantly reduced values of $v_m$, including  2800 m/s for ZrCoBi\cite{zhu2018discovery} and even lower values (2500-2800 m/s) for other lanthanide-based HHs\cite{huang2022discovery, ciesielski2020thermoelectric}.

Figure \ref{fig:structures}c shows that the speed of sound in ErNi$_x$Sb system can be even further reduced.  
Fully ionic ErSb has low speed of sound (\textit{ca.} 2400 m/s), and a linear relationship is found between the rock salt and half Heusler end members.  
Noting that the unit cell volume is nearly constant with $x$, the density $d$ of  ErNi$_x$Sb grows significantly with $x$ (see Tab. S1).  Extracting the bulk modulus ($B$), calculated as $B = v_{m}^2 d$, shows that the lattice stiffens with increasing $x$ (Tab. S1). To understand why nickel incorporation leads to stiffening, we consider the structural chemistry of half-Heusler materials; they are known to comprise ionic and covalent atomic interactions\cite{graf2011simple, bende2014covalence, zeier2016engineering}. The ionic part is primarily related to the rocksalt sublattice (Er-Sb for the studied system), while the covalent interactions stem from ionic zincblende sublattice (Ni-Sb in our case). The ionic compounds are expected to be softer than the covalent ones\cite{kraft2017influence, isotta2023elastic}. Hence, the incorporation of Ni and related change towards mixed ionic-covalent bonding can explain stiffening of the lattice in ErNi$_x$Sb. 
Such reduction in $v_m$ can contribute to suppression of thermal conductivity (Section \ref{sect:thermal}), which has been a long-lasting goal for the community studying HH thermoelectrics\cite{zeier2016engineering, quinn2021advances}. %An additional contribution to the modification in the speed of sound may be related to  its electronic properties, see Sect. 3.3.
Additionally, changes in electronic properties throughout ErNi$_x$Sb space can contribute further modifications to the speed of sound, see Sect. 3.3.

\begin{figure} [t]
\centering
  \includegraphics[width=1.005\columnwidth]{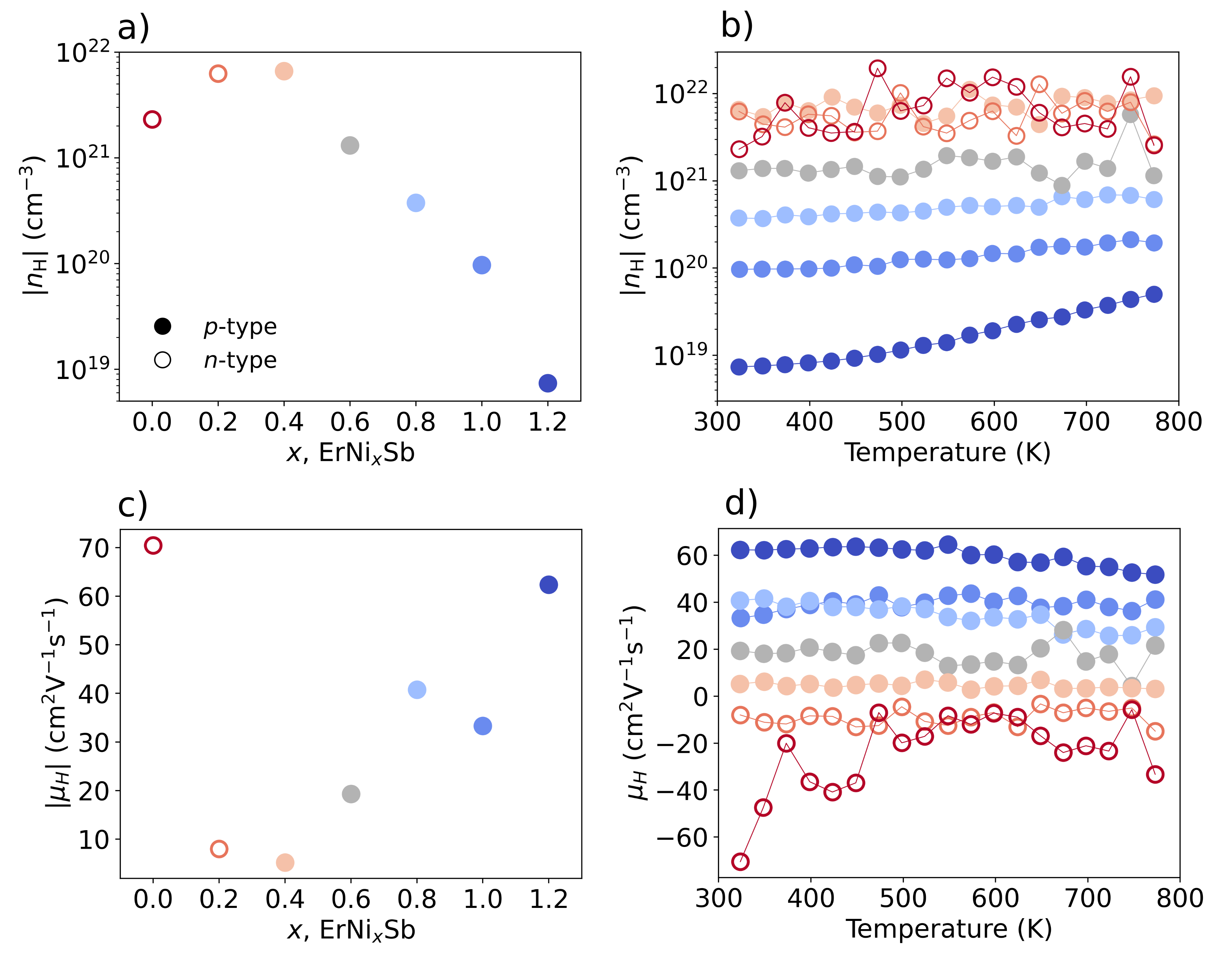}
  \caption{Hall data for ErNi$_x$Sb system. The top panels display carrier concentration (a) as a function of nickel content at room temperature, and (b) as temperature dependence. Values of $n_H$ increase 3 orders of magnitude with removal of nickel, which proves the metal-semiconductor transition. The bottom panels show charge mobility in analogous representation: (c) as a function of $x$ at RT, and (d) as a temperature variation. The most disordered alloys show decreased values of $\mu_H$.}
  \label{fig:Hall}
\end{figure}

\subsection{Electronic transport} 

Discussion of the electrical properties will start from the Hall data. Panels (a) and (b) of Fig. \ref{fig:Hall} display the carrier concentration ($n_H$) calculated within the single-band model. The sample with the highest Ni content (ErNi$_{1.2}$Sb) shows $p$-type conductivity, consistently with defect calculations\cite{chen2024ni}. Its carrier concentration is in range typical for semiconductors, \textit{ca.} 8$\times$10$^{18}$ carriers/cm$^3$, see Fig. \ref{fig:Hall}a. Temperature dependence $n_H$ for this sample shows thermal activation behavior, with bipolar onset at ca. 450 K, see Fig. \ref{fig:Hall}b. 
With removal of nickel in ErNi$_x$Sb, we observed metal-semiconductor transition. The $|n_H|$ values monotonically increase throughout 3 orders of magnitude as $x$ decays to 0.4. These results are consistent the band structure of the end-members, where ErNiSb has a band gap of \textit{ca.} 0.3 eV while ErSb has a negative gap of $\sim$1.5 eV\cite{lukoyanov2022ab}. For these $p$-type compositions, the temperature dependence flattens (Fig.\ref{fig:Hall}b), in line with expectations for degenerate semiconductors and metals.   

\begin{figure} [t]
\centering
  \includegraphics[width=\columnwidth]{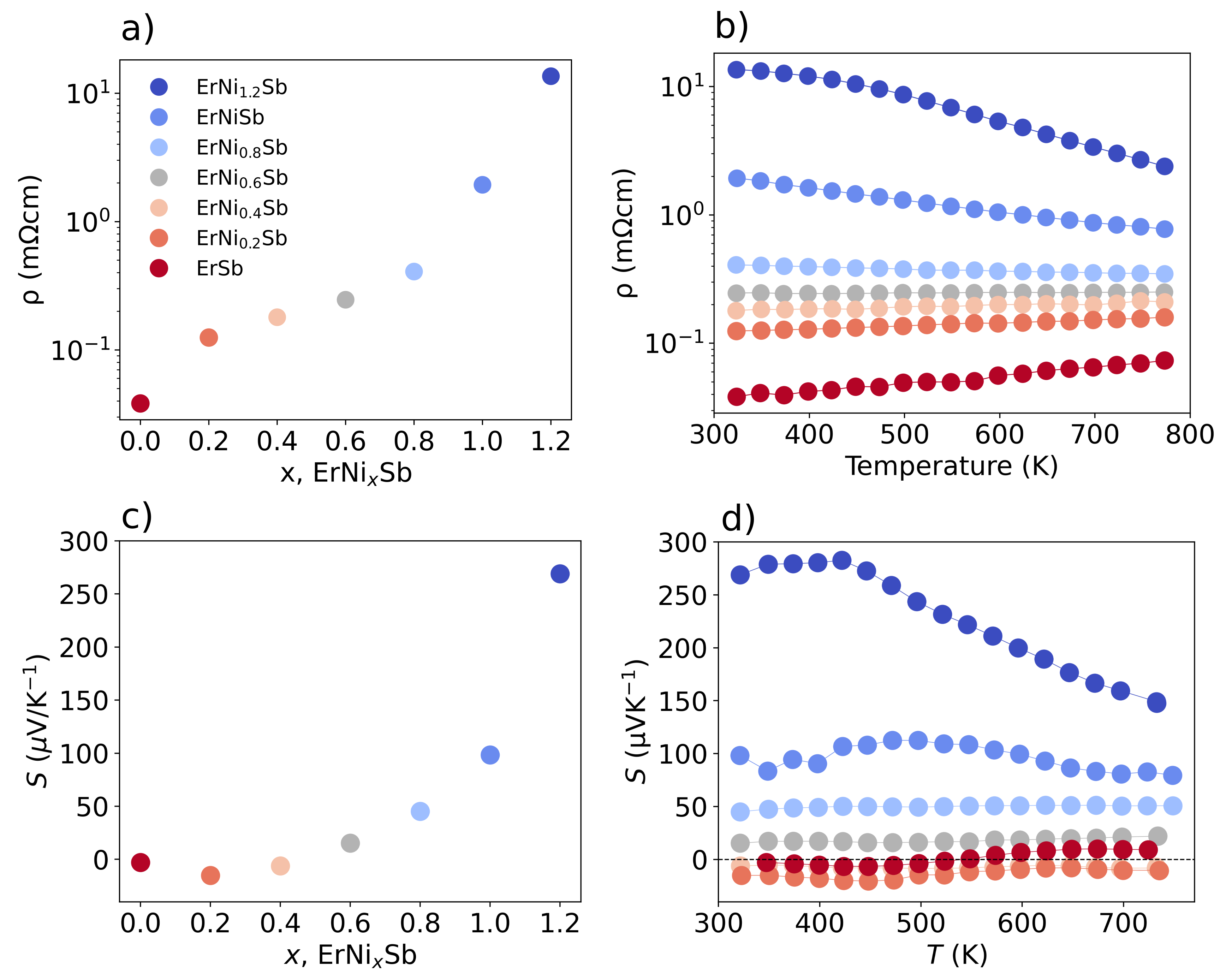}
  \caption{Electrical resistivity and Seebeck coefficient are showing metal-semiconductor transition in ErNi$_x$Sb system. The electrical resistivity is shown in panel (a) as a function of nickel content in and in panel (b) as temperature dependence. The bottom panels display thermopower as a function of composition (c), and temperature (d). Dashed line at $S$ = 0 is a guide to the eye.}
  \label{fig:r_S}
\end{figure}

The most Ni-poor specimens ($x$ = 0, 0.2) transition towards $n$-type behavior.  However, given the high conductivity of these samples (Fig. \ref{fig:r_S}), we expect that this $n$-type response is a reflection of bipolar transport.  The higher noise to signal ratio in the Hall measurements arises from a decreased magnitude of Hall voltage.  The increase in carrier concentration with removal of Ni can also be partially contributing to observed changes in sound velocity (Sect. 3.2), \textit{i.e.} we might be observing carrier-mediated lattice softening\cite{slade2021charge} in ErNi$_x$Sb.

The Hall mobility ($\mu_H$) at room temperature for the $x\geq 0.4$ compositions rises linearly with increasing nickel content (Fig.~\ref{fig:Hall}c). 
This trend can be understood as a result of linear changes in $V_{Ni}$ concentration and the increase in charge carrier concentration with smaller \textit{x}. At higher temperatures, mobility (Fig.~\ref{fig:Hall}d) found to be largely temperature-independent, indicating on dominance of point-defect scattering\cite{xie2014intrinsic}. 
For the two $n$-type compositions ($x$ = 0, 0.2), the mobility 
should be treated only as an approximation due to potentially multiband transport within the single parabolic band approximation used to estimate $n_H$ and $\mu_H$. The noticeable scatter in  mobility signal for Ni-poor samples is due to smaller magnitude of Hall constant, as discussed above for the carrier concentration.

Electrical resistivity ($\rho$) data is displayed in panels (a) and (b) of Fig. \ref{fig:r_S}. Consistent with the metal-semiconductor transition discussed previously, values of $\rho$ decrease by almost 3 orders of magnitude with removal of nickel from ErNi$_x$Sb. The temperature dependence of the resistivity also changes gradually from activation-like for ErNi$_{1.2}$Sb to metallic for ErSb.

Both the sign and magnitude of Seebeck coefficient (\textit{S}) for the ErNi$_x$Sb series (Fig. \ref{fig:r_S}d) are consistent with the Hall and resistivity results. 
The temperature of minority carrier activation in the Seebeck coefficient of ErNi$_{1.2}$Sb is similar to that observed in the Hall data (Fig. \ref{fig:Hall}b).
The highest observed thermopower is \textit{ca.} 265 $\mu$V/K for ErNi$_{1.2}$Sb at 450 K. The temperature dependencies for the Seebeck coefficient of ErNiSb and ErNi$_{0.8}$Sb are surprisingly weak. Effective masses calculated with SPB for samples in semiconducting regime ErNi$_{x}$Sb, $x$ = 0.6-1.2 are within the narrow range of 0.85-1.1$m_e$, consistent with previous reports for ErNiSb\cite{kawano2008substitution, ciesielski2020thermoelectric}. Such effective masses are lighter than the classical HH material ZrNiSn, $m_{eff}=2.8m_e$\cite{xie2014intrinsic}, and much lighter than the new generation of heavy band HH\cite{fu2015realizing, zhu2019discovery, zhu2018discovery} \textit{e.g.} FeNbSb, with $m_{eff}=6.8m_e$\cite{fu2015realizing}. The mobility follows the inverse trend to effective masses, being at room tempereature roughly 30 cm$^2$V$^{-1}$s$^{-1}$ for ZrNiSn\cite{xie2014intrinsic}, and \textit{ca.} 15 cm$^2$V$^{-1}$s$^{-1}$ or lower for the heavy band HH thermoelectrics\cite{fu2015realizing, zhu2019discovery, zhu2018discovery}.

To briefly sum up, comprehensive electrical characterization proven that preparation of rocksalt to HH solid solution is an excellent way to tune charge transport in wide range of behaviors.

\subsection{Thermal conductivity}
\label{sect:thermal}

\begin{figure} [t]
\centering
  \includegraphics[width=\columnwidth]{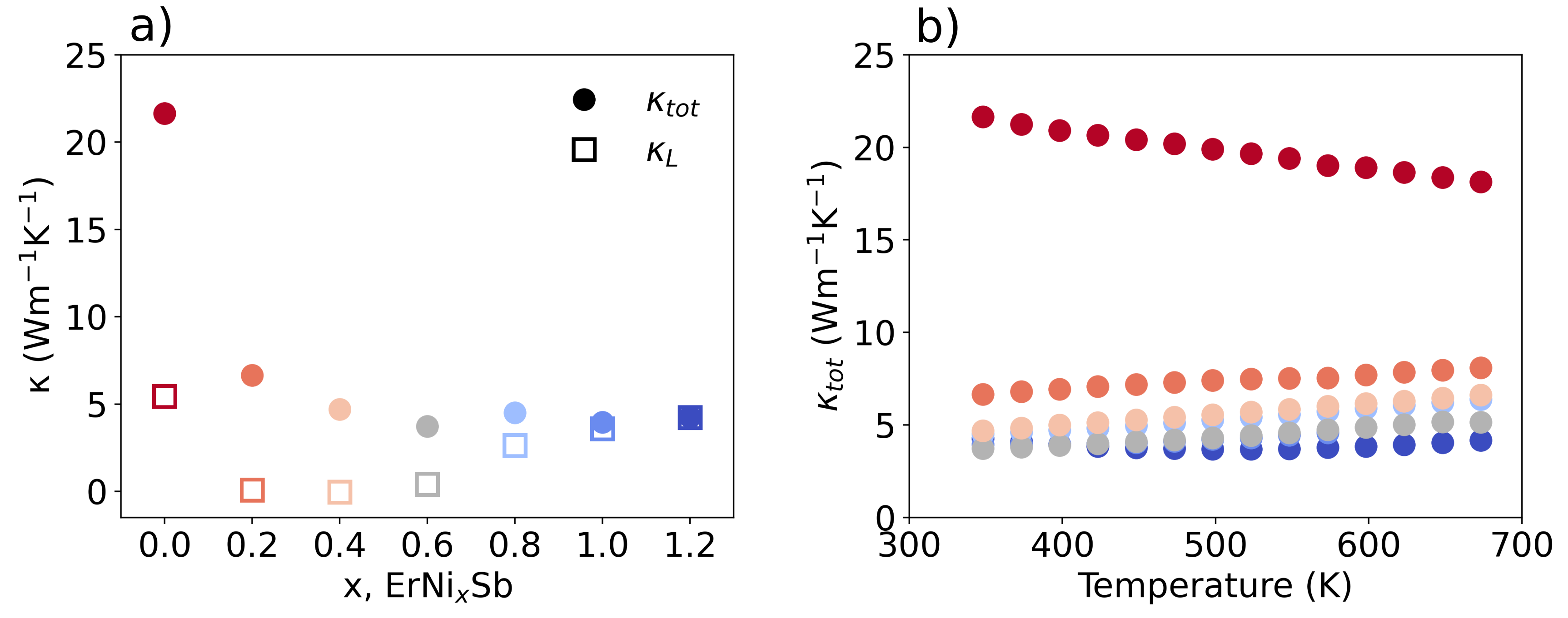}
  \caption{Disordering in ErNi$_x$Sb solid solution leads to suppresion  of phonon propagation. Panel (a) shows room temperature values of total and lattice thermal conductivity. Panel (b) displays temperature variation of total thermal conductivity.}
  \label{fig:thermal}
\end{figure}

Thermal conductivity ($\kappa_{tot}$) measurements are shown in Fig.~\ref{fig:thermal}a,~b. Values of $\kappa_{tot}$ are large for ErSb, consistent with expectation of for a good crystallographic order in its structure. Furthermore, very low $\rho$ of ErSb leads to a sizable electronic contribution to thermal transport. With the incorporation of Ni and the associate increase of $\rho$, the total thermal conductivity in ErNi$_x$Sb system settles around 5 Wm$^{-1}$K$^{-1}$ at RT. The temperature temperature variation of $\kappa_{tot}$ for ErNi$_{1.2}$Sb also shows onset of bipolar activation in middle temperature range (\textit{ca.} 500 K), similarly to Hall and Seebeck data.

To get more insight into mechanisms of thermal transport in the studied materials, we calculated the the electronic contribution to thermal transport ($\kappa_e$) with Franz-Wiedemann Law: 
$\kappa_{e}=LT\sigma$.
The equation $ L= 1.5 + \mathrm{exp}[\frac{-\lvert S \rvert}{116}]$
was used to compute the Lorenz number\cite{Wiedeman-Franz}. The lattice thermal conductivity ($\kappa_L$) is assumed to be the difference between total and electronic contribution to lattice thermal conductivity.  

Room temperature $\kappa_L$ is plotted using open symbols in Fig.~\ref{fig:thermal}a. For ErNiSb, the lattice thermal conductivity is around 4 Wm$^{-1}$K$^{-1}$. which is consistent with the previous measurement for ErNiSb showing range 4.5-5.5 Wm$^{-1}$K$^{-1}$, see Refs.\cite{ciesielski2020thermoelectric, kawano2008substitution}.
Thermal conductivity at RT for many other other prominent HH thermoelectrics in their pristine form are higher, see \textit{e.g.}: TaCoSn\cite{doi:10.1021/acsami.9b13603} (5.7 Wm$^{-1}$ K$^{-1}$),
NbCoSb\cite{xia_enhanced_2018} (7 Wm$^{-1}$ K$^{-1}$) 
ZrNiSn\cite{C4MH00142G} (9 Wm$^{-1}$K$^{-1}$), FeNbSb\cite{fu_realizing_2015} (18 Wm$^{-1}$ K$^{-1}$). Rare-earth based HH were generally described as having beneficial low $\kappa_L$ compared to other HH due to it is due to low speed of sound and intrinsic defects, especially on the late transition metals site\cite{ciesielski2020thermoelectric, huang2022discovery}.

ErSb, as a compound with no defects on the Ni site, shows the highest $\kappa_L$ at RT. Alloying between rocksalt and HH structures resulted in very significant decrease in $\kappa_L$ with respect to the parent compounds. For the samples ErNi$_x$Sb $x$ = 0.2, 0.4, 0.6, $\kappa_L$ at RT was below 0.6 Wm$^{-1}$K$^{-1}$. Here, it is necessary to underline, that the exact values of $\kappa_L$ for alloys should be treated with caution. These samples have more that 50\% of $\kappa_{tot}$ coming from the electronic contribution. Calculating proper Lorenz number can burdened with some error in such cases\cite{sawtelle2019temperature, shawon2024alloying, kumar1993experimental}. Employment of typical formula used for thermoelectrics\cite{Wiedeman-Franz} can sometimes even result in unphysical negative magnitude of $\kappa_L$\cite{shawon2024alloying}. Such problem also occurred in our data at elevated temperatures, see Fig. S1b. %\blue{Hence, the suppression of phonon transport in ErNi$_x$Sb should be treated only qualitatively.}

To quantitatively resolve the influence of disordering on $\kappa_L$ in the  rocksalt-HH alloy, we would need a compound with much lower electronic contribution to thermal transport. Preliminary attempts of subtle back-doping in ErNi$_{0.6}$Sb$_{1-y}$Te$_y$ failed, as the compound retained high electrical conductivity. To achieve this goal, a HH compound with a larger band gap could be used. 
For example, ScPtSb-ScSb system could be a potential candidate given the 0.6 eV\cite{winiarski2018thermoelectric} band gap.
%Potentially, a worthwhile playground for this idea can be ScPtSb-ScSb system ($E_g$ for ScPtSb $\approx$ 0.6 eV\cite{winiarski2018thermoelectric}).

The thermoelectric figure of merit (Fig. S1a) shows the highest values for the samples in semiconducting regime: ErNiSb and ErNi$_{1.2}$Sb. Tuning of the carrier concentration for the samples with the lowest $\kappa_L$ to achieve the highest possible performance was beyond the scope of this project. We anticipate that the opportunity of introduce severe disorder into HH lattice can be useful for other thermoelectrics from this chemical group, especially those that have rocksalt binary parent. 
Such continuous tuning of electronic and thermal properties can also be of interest for the broader community studying HH compounds from other perspectives, including topological properties and superconductivity\cite{graf2011simple, lin2010half, nakajima2015topological}.

\section{Conclusions}

In this work we demonstrated existence of complete solid solution between rocksalt ErSb and half-Heusler ErNiSb. Sound velocity measurements indicate, that the crystal lattice of ErNi$_x$Sb is gradually softening with the removal of nickel. The finding is in line with our understanding of changes in chemical bonding from ionic in ErSb to mixed ionic-covalent for ErNiSb. Carrier concentration, electrical resistivity and thermopower measurements demonstrated, that ErNi$_x$Sb undergoes a metal-semiconductor transition. The finding is in concert with the previous band structure calculations, in which ErSb is found to be a metal, while ErNiSb has a  0.3 eV band gap. The plethora of vacancies found within the solid solution turn out to be  efficient source of phonon scattering.
We anticipate that the approach to tuning transport can be used for other  materials from HH family, especially those based on rare-earth elements, which have been studied not only as thermoelectrics, but also unconventional superconductors and topological materials.

\section*{Conflicts of interest}
There are no conflicts to declare.

\section*{Acknowledgements}
The authors acknowledge support from NSF OAC 2118201.

\balance

\bibliography{rsc} 
\bibliographystyle{rsc} 

\end{document}